\documentclass[]{article}

\usepackage{graphicx}
\usepackage{dcolumn}
\usepackage{bm}
\usepackage[utf8]{inputenc}
\usepackage[T1]{fontenc}
\usepackage{mathptmx}
\usepackage{amsfonts}
\usepackage{bbm}
\usepackage[bottom]{footmisc}
\usepackage[utf8]{inputenc}
\usepackage{amsmath}
\usepackage{amsthm}

\fontsize{11}{16}\selectfont

\begin{document}

\title{The final version of a recent approach towards quantum foundation}

\author{Inge S. Helland, Department of Mathematics. University of Oslo\\ P.O. Box 1053, N-0316 Oslo, Norway\\ ingeh@math.uio.no\\ ORCID: 0000-0002-7136-873X }

\date{}

\maketitle

\begin{abstract}
In several articles, this author has advocated an alternative approach towards quantum foundation based upon a set of postulates, and based upon the notions of theoretical variables and of accessible theoretical variables.  It is shown in this article that this basis can be considerably simplified. In particular, the assumption that there exists an inaccessible variable $\phi$ such that all the accessible ones can be seen as functions of $\phi$, can be dropped from the basis, but it has other interesting consequences.The essential assumption is that there in the given context  exist two different maximal accessible variables, what Niels Bohr would have called two complementary variables. From this, the whole Hilbert space formalism may be derived. It is also discussed in some detail how this Hilbert space can be chosen. The resulting theory is a purely mathematical theory, but it leads to quantum mechanics by letting the variables be physical variables. Other applications of the main theory are also considered. The mathematical proofs are mostly deferred to the Appendix.
\end{abstract}

Keywords: accessible variables; complementary variables; Hilbert space formalism; inaccessible variables; quantum theory reconstruction; theoretical variables.

\section{Introduction}

In a number of recent articles, this author has sketched a completely new approach towards quantum foundation. The mathematical basis for this foundation is given in the articles Helland [5, 8], but this basis was not there in its final form. 

The fundamental notion of theoretical variables that may be accessible or inaccessible is very important, and this notion is shown to have applications also outside quantum mechanics, for instance in connection to statistical modelling (Helland, [9, 10]) and in psychology, exemplified by a new foundation of Quantum Decision Theory, see Helland [4]. This last application is consistent with Andrei Khrennikov's development of quantum-like models; see for instance Khrennikov [13], which points at numerous macroscopic consequences of quantum theory.

There are also wide discussions about interpretations of quantum mechanics in the literature. In my articles, I have advocated a general epistemic interpretation, which has QBism as a particular sub-interpretation. This will be further commented upon below.

The purpose of the present paper is to give a final mathematical foundation of my theory. In my earlier papers, a number of postulates were formulated, some of them rather obvious, but one postulate has been more difficult to motivate: In all my papers, I have assumed that there exists a basic inaccessible variable $\phi$ such that all the accessible ones can be seen as functions of $\phi$. In the mathematical developments below, I will here give a theory where this particular postulate can be dropped.

\section{The basis}

It is crucial to stress that the basic theory here is a purely mathematical theory. Once this theory has been laid down, various implications can be derived by giving interpretations of the mathematical concepts. One important implication is the foundation of quantum mechanics, another is the foundation of Quantum Decision Theory, and a third implication gives links to some statistical theory.

The basic notion is that of a theoretical variable, which is left to be undefined in the mathematical theory. The theoretical variables may or may not be accessible, again taken as an undefined notion. In this paper, I let the theoretical variables be real scalars, real vectors or real matrices, which is enough to give a rich theory. I only assume the following: If $\lambda$ is a theoretical variable, and $\theta = f(\lambda)$, a Borel-measurable function of $\lambda$, then $\theta$ is a theoretical variable. And if $\lambda$ is accessible, then $\theta$ is accessible.

Define a partial ordering among the theoretical variables, and also among the accessible ones, as follows: Say that $\theta \le \lambda$ if $\theta = f(\lambda)$, a Borel-measurable function of $\lambda$. If $f$ here is a bijective function, $\theta$ and $\lambda$ contain the same information, and we say that $\theta \sim \lambda$, $\theta$ and $\lambda$ are equivalent.

I postulate that there exist maximal accessible variables with respect to this partial ordering. More specifically, I assume: For any accessible theoretical variable $\zeta$, there exists a maximal accessible variable $\eta$ such that $\zeta \le \eta$. 

Furthermore, for some given maximal accessible variable $\theta$, I assume that there exists a transitive group $G$ acting on its range $\Omega_\theta$ such that $G$ has a trivial isotropy group and a left-invariant measure $\mu$. Given this, we can define the regular representation $U_L$ of $G$ by $U_L(g)f(\theta) = f(g^{-1}\theta)$ for $f \in L^2(\Omega_\theta, \mu)$.

Simple conditions that $G$ must satisfy in order that it shall have a left-invariant measure $\mu$, are given by Theorem 1 in Helland [5].

Using physical situations, it is easy to give examples where the above assumptions hold. One example is the position of a particle with a translation group. Another example may be the spin component of an electron in some direction, where the group is given by rotations of the Bloch sphere.

Altogether, these are weak assumptions on the theoretical variables and on the accessible theoretical variables. This basis is much simpler than taking as a point of departure that states are defined by normalized vectors in a complex Hilbert space. And it seems to be simpler than other reconstructions of quantum mechanics in the literature.

Later in the paper, I wll argue for a general epistemic interpretation of quantum mechanics. It is also of interest that this approach also has links to quantum field theory and to general relativity theory; see Helland and Parthasarathy [12].

\section{The main Theorems}

Given the basis above, a crucial assumption is that there in some given context exist two non-equivalent maximal accessible variables $\theta$ and $\eta$ with similar ranges, what Niels Bohr may have called two complementary variables.
\bigskip

\textbf{Theorem 1.}

\textit{Assume that in some given context, there exist two non-equivalent maximal accessible variables $\theta$ and $\eta$ such that}

\textit{(i). $\theta$ and $\eta$ have similar ranges, that is, there exists a bijective function $f_b$ between $\Omega_\theta$ and $\Omega_\eta$.}

\textit{(ii). There exists transitive group $G$ acting on its range $\Omega_\theta$ such that $G$ has a trivial isotropy group and a left-invariant measure $\mu$.}

\textit{Then there exists a Hilbert space $\mathcal{H}$, which can be taken as $ L^2 (\Omega_\theta, \mu )$, and there exist two symmetric operators $A^\theta$ and $A^\eta$ in $\mathcal{H}$ corresponding to $\theta$ and $\eta$.}
\bigskip

Using the basis of the previous Section, Theorem 1 is proved in the Appendix below. This proof is valid when $\theta$ takes at least 3 different values. The case with 2 values (the qubit case) is given a separate discussion in Helland [3], Chapter 4.

Under weak technical conditions given in Hall [2] and Helland [8], the two operators $A^\theta$ and $A^\eta$ will be self-adjoint. This is essential in order that the theorem shall form a basis for quantum theory. For self-adjoint operators the spectral theorem can be used.
\bigskip

\textbf{Proposition 1.}

\textit{If  $A^\theta$ and $A^\eta$ are self-adjoint, then to every accessible accessible variable $\zeta$ there corresponds a self-adjoint operator $A^\zeta$.}
\bigskip

\underline{Proof.}

By the basic assumptions, there exists a maximal accessible variable $\eta$ and a Borel-measurable funtion $f$ such that  $\zeta = f(\eta)$. This $\eta$ can be paired with the complementary variable $\theta$ of Theorem 1. The self-adjoint operator $A^\eta$ has a spectral decomposition
\begin{equation}
A^\eta = \int_{\sigma} x dE(x).
\label{xxx}
\end{equation}
where $\sigma$ is the spectrum of $A^\eta$ and $E$ is the spectral measure.

From this, we can define
\begin{equation}
A^\zeta = \int_{\sigma} f(x) dE(x).
\label{xxxx}
\end{equation}

It is easy to see that $A^\zeta$ is self-adjoint.

\qed
\bigskip

There is also a relationship between the two operators of Theorem 1.
\bigskip

\textbf{Theorem 2.}

\textit{There is a unitary operator $S$ in $\mathcal{H}$ such that}
\begin{equation}
A^\eta = S^{-1} A^\theta S.
\label{uu}
\end{equation}
\bigskip

The proof of Theorem 2 is also given in the Appendix.
\bigskip

Theorem 1 simplifies considerably when $\theta$ and $\eta$ take a finite number $r$ of values. Then the group $G$ is just a permutation group, and the operators $A^\theta$ and $A^\eta$ are trivially self-adjoint, since all finite-dimensional symmetric operators are self-adjoint. The egenvalues and eigenvectors of the operators have simple interpretations.
\bigskip

\textbf{Theorem 3.}

\textit{Assume that $\theta$ and $\eta$ both take $r$ values, and let $\zeta\le\eta$ be an arbitrary accessible variable.}

\textit{(i). The eigenvalues of $A^\zeta$ are the possible values of $\zeta$.}

\textit{(ii). The variable $\zeta$ is maximal as an accessible variable if and only if all eigenvalues of $A^\zeta$ are nondegenerate.}

\textit{(iii). If $\zeta$ is maximal, the eigenvectors of $A^\zeta$ can be interpreted as state vectors in the followng sense: They are in one-to-one correspondence wth questions `What is $\zeta$?' together with sharp answers $\zeta = c$.}

\textit{(iv). In the general case, the eigenspaces of $A^\zeta$ have a similar interpretation.}
\bigskip

Again the proofs are given in the Appendix.

Theorem 3 indicates important relations between the theory here and textbook quantum mechanics. In general, we can simply define pure states with specifications $\theta =c$, where $\theta$ is a maximal accessible variable; see Subsection 6.1 below. This represents a considerable conceptual simplification. The traditional state concept connected to an abstract unit vector in some Hilbert space, has caused much confusion, and has let to a number of incompatible interpretations.

Note again that, given the basis of Section 2, the only assumption that we need is that there exist two complementary variables in the context considered. By definition, two accessible variables are called complementary if they are non-equivalent, maximal accessible, and have similar ranges (the two ranges can be put in one-to-one correspondence). This makes Niels Bohr's complementarity concept precise.

\section{The Hilbert space}

The proof of Theorem 1 is given in the Appendix. In this proof we are asked to choose a function $f_0 \in L^2(\Omega_\theta, \mu)$ such that $f_0$ is a bijective function of $\theta$. Here, $\theta$ is one of the two complementary variables in Theorem 1, $\Omega_\theta$ is the range of $\theta$, $G$ is a transitive group with a trivial isotropy group acting on $\Omega_\theta$, and $\mu$ is the left-invariant measure associated with $G$. In this Section, the question of whether such a function $f_0$ always can be found, will be addressed. This is important in connection to the proof, since we have:
\bigskip

\textbf{Lemma 1.}

\textit{Define the functions $f_g$ by $f_g(\theta) = f_0(g^{-1}\theta)$. If $f_0$ is a bijective function on $\Omega_\theta$, then there is a one-to-one correspondence between $g$ and $f_g$ (; $g\in G$).}
\bigskip

\underline{Proof.}

It is clear that if $g_1 \ne g_2$, then $f_{g_1} \ne f_{g_2}$. Assume that $f_{g_1} \ne f_{g_2}$. Then for at least one $\theta =\theta_1$ we have that $f_0 (g_1^{-1}\theta_1)\ne f_0 (g_2^{-1}\theta_1)$. Since $f_0$ is bijective, this implies $g_1^{-1}\theta_1 \ne g_2^{-1}\theta_1$, hence $g_1\ne g_2$. \qed
\bigskip

The following result of this Section is of some interest: In general it not always possible to find such an $f_0$ as a realvalued function, but it can always be found as a complexvalued function. And, since it can be found, we can use the left regular representation $U=U_L$, defined by $U_L(g) f(\theta) = f(g^{-1}\theta)$, in the proof. The general requirement to $U$ is that the functions $U(g)f_0$ should be in one-to-one relation with $g\in G$, and hence, by transitivity, with $\theta\in \Omega_\theta$. By Lemma 1, this will be satisfied by $U=U_L$ if $f_0$ is a bijective function on $\Omega_\theta$.

Before giving the main result of this Section, it might be instructive to look upon some examples of a choice of the crucial function $f_0$:
\bigskip

\textit{a) $\Omega_\theta$ is finite} 

Here, $G$ is a permutation group, and any bijective $f_0$ can be used.

\textit{b) $\Omega_\theta = \mathbb{R}^1$, and $G$ is any transitive group with a trivial isotropy group and with a left invariant measure.}

Construction for the case $d\mu =dx$. ($G$ is the translation group): In principle, one can try here to use for instance $f_0(x) = \mathrm{min} (\mathrm{exp}(x), \mathrm{exp}(-x))$. Here the integral $\int f_0(x) dx$ converges, but the problem is that $f_0(\theta)$ is not a bijective function of $\theta$. Since the integral must converge both as $x\mapsto +\infty$ and as $x\mapsto -\infty$, it is impossible to choose $f_0$ as a monotone function, which it has to be to be bijective. We conclude that it is impossible to find a suitable realvalued function $f_0$ here.

But we can choose $f_0$ as a continuous complex function:
\begin{equation}
f_0(x) = \mathrm{min} (\mathrm{exp}(x), \mathrm{exp}(-2x)) +i\mathrm{min} (\mathrm{exp}(2x), \mathrm{exp}(-x)).
\label{f0}
\end{equation}

For $x\ge 0$, this is $f_0 (x) = \mathrm{exp}(-2x) +i\mathrm{exp}(-x)$, while for $x<0$ it is $f_0 (x) = \mathrm{exp}(x) +i\mathrm{exp}(2x)$. This is a bijective, continuous, integrable function of $x=\theta$.

Construction related to b)  in general: Note that if $G$ should be as required acting upon $\mathbb{R}^1$ , then for fixed $\theta_0$, we have that $g\theta_0$, a continuous bijective function of $g$, must be a decreasing or increasing function. ($g$ is in one-to-one correspondence with $\theta$.) Then $\mu$ must have a cumulative function which is decreasing or increasing. Write $d\mu = dF$ for an increasing or decreasing function $F$, and take $f_0(x) = f_{01} (F(x))$, where $f_{01}$ is the $f_0$ from the previous point.

\textit{c) $\Omega_\theta = (0, \infty)$, and $G$ is the multiplication group.}

Introduce $\zeta = \mathrm{ln}(\theta)$. Then $G$ acts as a translation group on $\zeta$, and we can use b).

\textit{d) $\Omega_\theta$ consists of vectors of the form $\lambda \mathbf{d}$, where $\lambda$ is a  scalar, $\mathbf{d}$ a unit vector, $G$ acting on $\lambda$ is the multiplication group, and $G$ acting on $\mathbf{d}$ is the joint rotation group. }

Take $\zeta = \mathrm{ln}(\lambda)$. Let $f_{01}$ be the function (\ref{f0}) with $x$ replaced by $\zeta$, and define now $f_0 (\lambda \mathbf{d}) = f_{01}(\zeta)h(\mathbf{d})$, where $h$ is a bijective integrable function.

\textit{d) $\Omega_\theta$ consists of matrices of the form $(\lambda_1 \mathbf{d}_1, ... ,\lambda_r \mathbf{d}_r)$, where the $\lambda_i$'s are scalars, the $\mathbf{d}_i$'s are unit vectors, $G$ acting on the $\lambda_i$'s is the multiplication group, and $G$ acting on the $\mathbf{d}$'s is the joint rotation group.}

 Let $\zeta_j = \mathrm{ln}(\lambda_j)$ $(j=1,...,r)$. This gives a total function $f_0$ defined by $f_0(\lambda_1 \mathbf{d}_1, ... ,\lambda_r \mathbf{d}_r) ) = f_{01} (\zeta_1)\cdot ...\cdot f_{01}(\zeta_r) h_0(\mathbf{d}_1, ... ,\mathbf{d}_r)$, where $f_{01} (x)$ is given by (\ref{f0}), and $h_0$ is an integrable bijective function.

This case was needed in Helland [10].
\bigskip

It is now easy to prove a general theorem on the construction of $f_0$, and thus on the Hilbert space construction, atleast in the scalar case. It is of some interest to see when the Hilbert space is real, and when it must be complex. It is well known that quantum mechanics on real Hilbert spaces have different properties than quantum mechanics on complex Hilbert spaces; see for instance Stueckelberg [14].
\bigskip

\textbf{Theorem 4.} 

\textit{Let $\theta$ be a realvalued accessible theoretical variable, and define  $\Omega_\theta$, $G$ and $\mu$ as above. Then $f_0 \in L^2(\Omega_\theta, \mu)$ can always be found as a complexvalued function, the unitary representation $U = U_L$ can be used in the proof of the main theorem, and the resulting Hilbert space can be taken to be a $\mathcal{H} =  L^2(\Omega_\theta, \mu)$.}

\textit{ a) If  $\Omega_\theta$ is a set that is bounded as $\theta \mapsto -\infty$ by a value $\theta_1$, and $\mu (B)$ is finite when $B$ is an interval containing $\theta_1$, then $f_0$ can be found as a realvalued function, and $\mathcal{H}$ is based upon real numbers.}

\textit{b) If $\Omega_\theta$ is unbounded both as  $\theta \mapsto +\infty$ and  $\theta \mapsto -\infty$, then it is impossible to choose $f_0$ to be realvalued, and $\mathcal{H}$ must be based upon complex numbers.}
\[\]

\underline{Proof.}

 I will start by proving a). Let $\Omega_\theta$ be bounded below by some $\theta_1$. Then choose $f_0$ as a monotonically decreasing function for $\theta \ge \theta_1$ with finite $f_0(\theta_1)$. Then $f_0$ is a bijective function of $\theta$. By letting $f_0(\theta)$ decrease sufficiently fast towards $0$ as $\theta\mapsto\infty$, we may assume for any $\mu$ that $f_0 \in L^2(\Omega_\theta, \mu)$.
 
 Now it is easy to construct a complex function $f_0$ for the case where $\Omega_\theta$ is unbounded in both directions: Let $f$ be the function defined in the previous paragraph for $\theta$ larger or equal to some $\theta_1$, which without loss of generality can be taken to $\theta_1 = 0$. Define $f_0 (x) = f(x) + if(2x)$ for $x\ge 0$ and $f_0(x) = f(-2x)+if(-x)$ for $x\le 0$. Then $f_0$ is bijective and belongs to $L^2(\Omega_\theta, \mu)$ for any $\mu$.
 
 It is clear that no realvalued $f_0$ can do this job when $\Omega_\theta$ is unbounded in both direction. Such an $f_0$ has to be monotonically decreasing for $x\ge 0$, and should it be bijective, it must also be monotonically decreasing for $x\le 0$. But then it cannot tend to $0$ as $x=\theta$ tends to $-\infty$, and, if $\mu$ is nontrivial for large negative $\theta$, it cannot belong to $L^2(\Omega_\theta ,\mu )$.
 
\qed

\section{Concequences of assumptions on a basic inaccessible variable.}

It is not necessary for the derivation of the Hilbert space formalism, but in some cases, we can make the following assumption.
\bigskip

\textbf{Assumption 1.}

\textit{There exists an inaccessible variable $\phi$ with the property that all accessible variables can be seen as functions of $\phi$. There is a group $K$ acting on the range $\Omega_\phi$.}
\bigskip

A physical example is given by the spin components of an electron. Let $\phi$ be a random vector from the origin to the surface of tthe Bloch sphere. Let $K$ be the rotation group. Then the spin component in direction $a$ may be modelled by
\begin{equation}
\theta^a = \mathrm{sign}(\mathrm{cos}(\phi, a)).
\label{spin}
\end{equation}

Another example is the following: Let $\theta$ be the position and $\eta$ the momentum of a particle. Take $\phi$ as the vector $(\theta, \eta )$, and let $K$ be the Heisenberg-Weyl group.
\bigskip

\textbf{Definition 1.}

\textit{Two maximal accessible variables $\theta$ and $\eta$ are said to be related (relative to $\phi$) if there is a transformation $k$ in $\Omega_\phi$ and a function $f$ such that}
\begin{equation}
\theta = f(\phi),\ \ \ \eta = f(k\phi).
\label{related}
\end{equation}

\textbf{Lemma 2.}

\textit{ Assume that there is a group $K$ acting upon $\Omega_\phi$. Then every pair $\theta$ and $\eta$ of non-equivalent maximal accessible variables with similar range spaces have the following property: There exists an accessible variable $\xi$ which is a bijective function of $\eta$ such that $\xi$ and $\theta$ are related.}
\bigskip

\underline{Proof}

See the proof of Proposition 1 in Helland [8]. \qed
\bigskip

In many cases, one can take $\xi=\theta$, so that $\eta$ and $\theta$ are directly  related. This is the case for electron spin components, where we can let $K$ be rotations of the Bloch sphere.

By Theorem 2, if $\theta$ and $\eta$ satisfy the conditions of Theorem 1, then the corresponding operators satisfy the relation $A^\eta = S^{-1}A^\theta S$ for a unitary operator $S$, what we will call a similarity relation. We also have the following result:
\bigskip

\textbf{Theorem 5.}

\textit{If $\theta$ and $\eta$ are related through a transformation $k$, then $A^\theta$ and $A^\eta$ satisfy a similarity relation with $S=W(k)$. On the other hand, if $A^\theta$ and $A^\eta$ satisfy a similarity relation, then $\theta$ and $\eta$ are related relative to some $\phi$ such that $\theta$ and $\eta$ are functions of $\phi$.}
\bigskip

\underline{Proof}

See the proofs of Theorem 2 and Theorem 6 in Helland [8].    \qed

\section{Applications of the mathematical theory.}

\subsection{Quantum mechanics.}

Note that the theory so far has been a purely mathematical theory, where the notions of theoretical variables and accessible theoretical varables are left undefined. But now we can reconstruct quantum mechanics by interpreting these variables as physical variables. Two simple examples of pairs of complementary variables are: 1) Take $\theta$ as position and $\eta$ as momentum of a single particle. 2) Take $\theta$ and $\eta$ as spin components of an electron in two given directions.

The theory gives symmetric/ self-adjoint operators corresponding to all accessible theoretical variables. The natural state vectors are eigenvectors of these operators. From this, I propose the following version of quantum theory: As state vectors we only include vectors in the Hilbert space that are eigenvectors of a meaningful physical operator. 

Consider first the case where $\theta$ and $\eta$ take a finite number $r$ of values. Then the operators are just $r$-dimensional complex matrices. If $\theta$ is maximal accessible, the eigenvectors of the corresponding operator $A^\theta$ are non.degenerate, and the set of eigenvalues of $A^\theta$ are equal to the possible values of $\theta$. It is then natural to call a statement $\theta = c$ a state of the underlying physical system. Depending upon the situation, this statement/ state can be associated with an observer or with a group of communicating observers.

Note the important result that these states, as $\theta$ is allowed to vary, are in one-to-one correspondence with the normalized eigenvectors of the corresponding operators $A^\theta$. We can thus follow the ordinary textbook-foundation of quantum mechanics and call complex unit vectors for states.

These are pure states. Mixed states can be defined by probability distributions over the values of one maximal accessible variabl $\theta$. Now the pure states can also be characterized by one-dimensional projection matrices $P_k$, so the mixed states are equivalent to matrices $\sum_k P(\theta = c_k)P_k$, density matrices.

Now turn to the continous accessible variables $\theta$. Since these can always be approximated by finite-valued variables, we get mixed states of the form $\int_\sigma f_\theta (x) dE_\theta (x)$, where $f_\theta (x)$ is the density of a probability distribution of $\theta$, $\sigma$ is the spectrum of $A^\theta$, and $\{E_\theta \}$ is the spectral distribution of $A^\theta$. 

These density operators have a very simple interpretation: They are given by a probability disytribotion of an accessible variable $\theta$. One can start by maximal accessible variables, but show that the same interpretation is valid for any accessible variable. Again, depending upon the situation, the probability distribution can be associated with some human or with a communicating group of humans.

Note that all this starts with a premise: The allowable state vactors in traditional quantum quantum mechanics are always eigenvectors of some physically meaningful operator. 

This may be in conflict with the general superposition principle that is usual to assume, but on the other hand, it gives a version of quantum theory where for instance the paradox of Schr\"{o}dinger's cat disappears; see Helland and Parthasarathy [12]. I will discuss the superposition principle in general elsewhere.

As an example, an entangled state, the singlet state vector of the Bell experiment, is an eigenvector for the operator corresponding to the dot product of the two spin vectors; see Susskind and Friedman [15]. All vectors orthogonal to the singlet vector are also eigenvectors of the same operator. The singlet state can be shown to correspond to the case where the two spin vectors are opposite; see Subsection 6.2.1 in Gill et al. [17].

In general, superpositions of the following form are allowed, where I for simplicity limit myself to the finite-valued case: Let $\{|a_i\rangle\}$ be the normalized eigenvectors of an operator $A^a$, and let $|b\rangle$ be an arbitrary eigenvector of another operator $A^b$. Then $\sum |a_i\rangle\langle a_i| = I$, and
\begin{equation}
|b\rangle = \sum |a_i\rangle\langle a_i| |b\rangle = \sum \langle a_i |b\rangle |a_i\rangle.
\label{uuu}
\end{equation}

What is lacking in the above discussion, is a treatment of the Born rule, giving probabilities in quantum mechanics. In Helland [10] it is shown that the Born rule follows under an assumption that is closely related to the likelihood principle of statistics. In the Bayesian case, it is enough to assume that the unconditional probabilities for each state exist.

\subsection{The epistemic interpretation}

Taking the example of Wigner's friend as a point of departure, it is natural to couple the state vectors to some person $C$. This is also in agreement with Herv\'{e} Zwirn's convivial solipsism (Zwirn [16]), which is proposed in order to solve the measurement problem.

I will propose a generalization of this: The state vectors of quantum mechanics are associated with a single person or with a group of communicating persons. The group is assumed to be able to communicate about everything that is related to the relevant theoretical variables. Assume an interpretation of quantum mechanics where the state vectors describe the knowledge that $C$ (or the group) has about the world, not directly a theory of the world itself.

This is what I will call the general epistemic interpretation. A further discussion of this interpretation and the relationship to other interpretations is given in Helland [5,6]. A sub-interpretation of the general epistemic interpretation is QBism, see Caves et al.[1] and references there.

In very many cases, the assumed group of people may in principle consist of all persons in the world. Then the actual state vector has some objectivity property connected to it, and we may say that we have a link to an ontological interpretation of quantum mechanics.

\subsection{Quantum Decision Theory}

Let the person $C$ be in a situation where he has the choice between a set of actions $\{a_x\}$. In Helland [4] this set was supposed to be finite, but by using the theory of the present paper, it can also be infinite. Define a decision variable $\theta$ to be equal to the index $x$ if the action $a_x$ is to be chosen.

Let the decision variable be maximal if $C$ is just able to carry our the decision: If one more action had been in the set, he would have been unable to take the decision.

In some cases, $C$ would have in mind two different such decision processes. Then the result of Theorem 1 will apply, and we have a foundation of Quantum Decision Theory.

\subsection {Links to statistical theory}

In this interpretation, we may let the theoretical variables be statistical parameters.

In very many case in applied statistics, the natural parameter space is too large compared with the data that are available. Then a parameter reduction may be called for. In Helland [10] two such parameter reductions are compared, using essentially the situation described in Theorem 1.

Another application of Theorem 1 is described in Helland [9]. Here, two experiments are done, the first focuses on a subparameter $\theta$, the other with another subparameter $\eta$. It argued that, if both these subparameters are maximal, then a prior for the second experiment should be taken as a quantum probability.

\section{Conclusion}

For further discussions related to this approach, see the references below. In particular, the Born rule and the quantum probabilities are derived from two additional postulates in Helland [3, 7].

The purpose of the present article has been to show that this approach towards quantum theory may be developed from simple assumptions by a completely rigorous mathematical theory. I will claim after this that quantum mechanics may be derived from an intuitive set of assumptions: The hypothesis that there exist two non-equivalent complementary variables, two accessible theoretical variables that are maximal as accessible variables, and the rest is a fairly intuitive basis.

Note again that the pure states can be denoted as $\theta = c$, where $\theta$ is a maximal accessible variable, and that mixed states in general can be characerized by some classical probability distribution over a maximal accessible variable.

\section*{Acknowledgments}

I am grateful to Trygve Alm\o y, Solve S\ae b\o , Richard Gill and Bart Jongejan for discussions. In particular, a recent  discussion with Richard Gill has motivated me to write this article.

\section*{References}

$\ \ \ \ $ [1] Caves,, C.M., Fuchs, C.A., and Schack, B. (2002). Quantum probabilities as Bayesian probabilities. \textit{Physical Review} A \textbf{65}, 022305.

[2] Hall, B.C. (2013) \textit{Quantum Theory for Mathematicianx.} Springer, Berlin.

[3] Helland, I.S. (2021) \textit{Epistemic Processes. A Basis for Statistics and Quantum Theory.} 2. Edition. Springer Nature, Cham, Switzerland.

[4] Helland, I.S. (2023). A simple quantum model linked to decisions. \textit{Foundations of Physics} \textbf{53}, 12.

[5] Helland, I.S. (2024a). An alternative foundation of quantum mechanics. arXiv: 2305.06727 [quant-ph]. \textit{Foundations of Physics} \textbf{54}, 3.
 
 [6] Helland, I.S. (2024b). A new approach towards quantum foundation and some consequences. arXiv: 2403.09224 [quant-ph]. \textit{Academia Quantum} \textbf{1}, 7282.
 
 [7] Helland, I.S, (2024c). On probabilities in quantum mechanics. \textit{APL Quantum} \textbf{1}, 036116.

[8] Helland, I.S. (2025a). Some mathematical issues regarding a new approach towards quantum foundation. arXiv: 2411.13113 [quant-ph]. \textit{Journal of Mathematical Physics} \textbf{66}, 092103.

[9] Helland, I.S (2025b). Quantum probability for statisticians: Some new ideas. \textit{ Methodology and Computing in Applied Probability} \textbf{27} (84), 1-24.
  
[10] Helland, I.S. (2026a). On optimal linear prediction. Discussion paper. \textit{Scandinavian Journal of Statistics} \textbf{53} (1), 16-32.

[11] Helland, I.S. (2026b). On the Born rule in a new quantum approach, ArXiv 2609.24269 [quant-ph].

[12] Helland, I.S. and Parthasarathy, H. (2024). \textit{Theoretical Variables, Quantum Theory, Relativistic Quantum Field Theory, and Quantum Gravity.} Manakin Press, New Dehli.

[13] Khrennikov, A. (2010). \textit{Ubiquitous Quantum Systems. From Psychology to Finance.} Springer, Berlin.

[14] Stueckelberg, E.C.G. (1960). Quantum theory in real Hilbert space. \textit{Helvetica Physical Acta} \textbf{33} (727) 458.

[15] Susskind, L. and Friedman, A. (2014). \textit{Quantum Mechanics. The Theoretical Minimum.} Penguin Books, New York.

[16] Zwirn, H. (2016). The measurement problem: Decoherence and convivial solipsism. \textit{Foundations of Physics} \textbf{46}, 635-667.

[17] Gill, R., Helland, I.S. and Jongejan, B. (2026).Three ways to find comfort with the Bell proof and the results of the Bell experiments. ArXiv 2605.13154 [quant-ph].

\section*{Appendix: Proofs of the main Theorems.}

\textbf{Proof of Theorem 1.}

Let $\phi = (\theta, \eta)$. I will define a group $N$ acting on $\phi$, and a representation $W$ of this group which is irreducible. This will be used to construct the operators $A^\theta$ and $A^\eta$.

First the construction of the group $N$: For $g\in G$, define $g(\theta,\eta) =(g\theta, \eta)$. Let $G^1$ be an independent copy of $G$, and let $H$ be the group acting on $\eta$ defined by $h\eta = f_b (g^1 \theta)$ when $\eta = f_b(\theta)$, and then $h(\theta,\eta)=(\theta,h\eta)$. Finally, let $j(\theta,\eta) = (\eta,\theta)$. Then define the group $N$ as the group generated by $G, H$ and the element $j$ as acting upon $\phi$. One can let $G$ act on $\eta$ by defining $g\eta =f_b(g\theta)$; similarly one can let $H$ act on $\theta$.

Note that this group is non-abelian: $jg(\theta,\eta) = (\eta, g\theta)$, while $gj(\theta,\eta) =(g\eta, \theta)$. Since $G$ and $H$ are transitive on their components, and since through $j$ one can choose for a group element of $N$ to act first arbitrarily on the first component and then arbitrarily on the second component, $N$ is transitive on $\phi$. Also, $N$ has a trivial isotropy group.

Consider $\Omega_\theta$, the group $G$ acting on $\Omega_\theta$, and the left regular representation $U = U_L$ of $G$ defined by $U(g)f(\theta) = f(g^{-1}\theta)$ for $f\in \mathcal{H} = L^2(\Omega_\theta,\mu)$. In Section 4 it was proved that we can find $f_0\in\mathcal{H}$ such that $U(g)f_0$ is in one-to-one correspondence with $g$ as $g$ varies over $G$. In this article, I will for simplicity assume that $U_L$ gives an irreducible representation of $G$ on $L^2 (\Omega_\theta .\mu )$. A more rigorous theory on this point would require the introduction of a rigged Hilbert space.Also, I will use Dirac's notation for states even in the case of a continuous variable $\theta$.

For each element $g\in G$ there is an element $h =j g j\in H$ and vice versa. Note that $j\cdot j=e$, the unit element. Let $U(j)=J$ be some unitary operator on $\mathcal{H}$ such that $J\cdot J =I$. Then for the representation $U (\cdot)$ of the group corresponding to $G$, there is a representation $V (\cdot)$ of the group corresponding to $H$ given by $V (j gj)=J U (g)J$. These representations are acting on the same Hilbert space $\mathcal{H}$, and they are equivalent in the concrete sense that the groups of operators $\{U(g)\}$ and $\{V(h)\}$ are isomorphic.

Since $U(g)f_0$ is in one-to-one correspondence with $g$, and hence by transitivity with $\theta$, we can write $|\theta\rangle = U(g)|\theta_0\rangle$, where the ket vector $|\theta_0\rangle$ is given by the function $f_0 \in \mathcal{H} = L^2 (\Omega_\theta, \mu)$. Similarly, we can write $|\eta\rangle = V(h)|\eta_0\rangle$.

Note that $J$ must satisfy $J U(j g j)=U (g)J$.  By Schur's Lemma, this demands $J$ to be an isomorphism or the zero operator if the representation $U(\cdot )$ was irreducible, which it is not in general. In the reducible case a non-trivial operator $J$ exists, however:

In such a case there exists at least one proper invariant subrepresentation $U_0$ acting on some vector space $\mathcal{H}_0$, a proper subspace of $\mathcal{H}$, and another proper invariant subrepresentation $U'_0$ acting on an orthogonal vector space $\mathcal{H}'_0$. Fix $|v_0\rangle \in \mathcal{H}_0$ and $|v'_0\rangle \in \mathcal{H}'_0$, and then define $J|v_0\rangle=|v'_0\rangle$, $J|v'_0\rangle=|v_0\rangle$ and if necessary $J|v\rangle=|v\rangle$ for any $|v\rangle\in \mathcal{H}$ which is orthogonal to $|v_0 \rangle$ and $|v'_0 \rangle$.

Now we can define a representation $W(\cdot)$ of the full group $N$ acting on $\phi =(\theta, \eta)$ in the natural way: $W(g)=U(g)$ for $g\in G$, $W(h)=V(h)$ for $h\in H$, $W(j)=J$, and then on products from this.
\bigskip

If $U$ is irreducible, then also $V$ is an irreducible representation of $H$, and we can define operators $A^\theta$ corresponding to $\theta$ and $A^\eta$ corresponding to $\eta$ by
\begin{equation} 
A^\theta = \int \theta |\theta\rangle\langle\theta| d\mu (\theta);\ \ \ A^\eta = \int \eta |\eta\rangle\langle\eta| d\mu (\eta).
\label{xx}
\end{equation}

By using Schur's lemma, we can show in this case that $\mu$ can be normalized such that
\begin{equation}
 \int  |\theta\rangle\langle\theta| d\mu (\theta) =I. 
 \label{x}
 \end{equation}

 Hence, these operators have the desirable properties:

(i) It $\theta = c$, then $A^\theta = cI$.

(ii) If $\theta$ is real-valued, then $A^\theta$ is symmetric.

(iii) The change of basis through a unitary transformation is straightforward.

If $U$ is reducible, we need to show that the representation $W$ of $N$ constructed above is irreducible.
\bigskip

\textbf{Lemma A1.} 

\textit{$W(\cdot)$ as defined above is irreducible.}
\bigskip

\underline{Proof.}

Assume that $W(\cdot)$ is reducible, which implies that both $U(\cdot)$ and $V(\cdot)$ are reducible, i.e., can be defined on a proper sub-space $\mathcal{H}_0\subset\mathcal{H}$, and that $J=W(j)$ also can be defined on this sub-space. Let $R(\cdot)$ be the representation $U(\cdot)$ of $G$ restricted to vectors $|u\rangle$ in $\mathcal{H}$ orthogonal to $\mathcal{H}_0$. Fix some vector $|u_0\rangle$ in this orthogonal space; then consider the coherent vectors in this space given by $R(g)|u_0\rangle$. Note that the vectors orthogonal to $\mathcal{H}_0$ together with the vectors in $\mathcal{H}_0$ span $\mathcal{H}$, and the vectors $U(g)|u_0\rangle$ in $\mathcal{H}$ are in one-to-one correspondence with $\theta$. Then the vectors $R(g)|u_0\rangle$. are in one-to-one correspondence with a subvariable $\theta^1$. And define the representation $S(\cdot)$ of $H$ by $S(jgj) =R(g)$ and vectors $S(h)|v_0\rangle$, where $|v_0\rangle$ is a fixed vector of $\mathcal{H}$, orthogonal to $\mathcal{H}_0$. These are in one-to-one correspondence with a subparameter $\eta^1$ of $\eta$.

Fix $\theta_0 \in \Omega_\theta$. Given a value $\theta$, there is a unique element $g_\theta \in G$ such that $\theta = g_\theta \theta_0$. (It is assumed that the isotropy group of $G$ is trivial.)

From this look at the vectors $S(jg_\theta j)|v_0\rangle$. By what has been said above, these correspond to unique values $\eta^1$, which are determined by $g_\theta$, and hence by $\theta$. But this means that a specification of $\theta$ leads to a new accessible vector $(\theta, \eta^1)$, contrary to the assumption that $\theta$ is maximal as an accessible variable.  Thus $W(\cdot)$ cannot be reducible.

\qed

\bigskip

This lemma shows that there are group actions $n\in N$ acting on $\phi = (\theta, \eta)$ and an irreducible representation $W(\cdot)$ of $N$ on the Hilbert space $\mathcal{H}$. Hence, the identity (\ref{x}) holds if $G$ is replaced by $N$, and the coherent states by $|v_n\rangle = W(n)|v_0\rangle$:

\begin{equation}
\int |v_n\rangle\langle v_n |\mu (dn)=I,
\label{u4}
\end{equation}
where $\mu$ is some suitably normalized left-invariant measure on $N$, and $|v_0\rangle$ is some fixed vector in $\mathcal{H}$. (Since $G$ and $H$ have left-invariant measures $\mu$ on $\Omega_\theta$ and on $\Omega_\eta$, respectively, there is also a left-invariant measure of $N$ on $\phi$, a measure that I also call $\mu$.)
\bigskip

 \textbf{Lemma A2.} 
 
 \textit{There is a function $f_\theta$ of $n$ such that $\theta=f_\theta (n)$, and a function $f_\eta$ of $n$ such that $\eta=f_\eta (n)$.} 
 \bigskip
 
 \underline{Proof.}
 
 Consider a transformation $n$ transforming $\phi_0 =(\theta_0 ,\eta_0 )$ into $\phi_1 =(\theta_1, \eta_1  )$. There is then a unique $g$ transforming $\theta_0$ into $\theta_1$, and a unique $h$ transforming $\eta_0$ into $\eta_1$. Since the groups $G$ and $H$ are assumed to be transitive and with a trivial isotropy group, the group elements $g$ and $h$ correspond to unique variable elements $\theta$ and $\eta$. These are then determined by $n$.
 
 \qed
 
 \bigskip

 We are now ready to define operators corresponding to $\theta$ and $\eta$:

\begin{equation}
A^\theta=\int f_\theta (n)|v_n\rangle\langle v_n |\mu (dn),
\label{u5}
\end{equation}

\begin{equation}
A^\eta=\int f_\eta (n)|v_n\rangle\langle v_n |\mu (dn).
\label{u6}
\end{equation}

It is clear that these operators are symmetric when $\theta$ and $\eta$ are real-valued variables. Under some weak technical assumptions they will be self-adjoint/ Hermitian.  Also, if $\theta=c$, then $A^\theta$ is $c$ times the identity. For this, the left-invariant measure $\mu$ is normalized (using Schur's lemma) such that
\begin{equation}
\int |v_n\rangle\langle v_n |\mu (dn) = I.
\label{uu}
\end{equation}

\bigskip

 \textbf{Proof of Theorem 2.} 

If $s$ is any transformation in $N$, and $W(\cdot)$ is the representation of $N$ used in the above proof, we have

\begin{equation}
W(s^{-1})A^\theta W(s)=\int f_\theta (sn)|v_n\rangle\langle v_n |\mu (dn),
\label{u7}
\end{equation}
\bigskip

\underline{Proof.}

\begin{equation}
W(s^{-1})A^\theta W(s) = \int f_\theta (n) W(s^{-1}n)|v_0\rangle\langle v_0| W(s^{-1}n)^{-1} \mu(dn).
\label{ux}
\end{equation}
Change the variable from $s^{-1}n$ to $n$ and use the left-invariance of $\mu$.
\qed
\bigskip

Consider an application of this: The statement of Theorem 2 follows from the fact that the transfomation $j$ acts on $\phi=(\theta, \eta)$ and induces a transformation $s(j)$ on the group $N$. Take $s=s(j)$ and $S=W(s(j))$ in (\ref{u7}). 

\bigskip

\textbf{Proof of Theorem 3.}

Consider the case where the maximal accessible variables as in Theorem 3 take a finite number of values. Note that the construction in Proposition 1 of an operator corresponding to a variable can be made for any maximal accessible variable $\zeta$. If $\zeta$ is not maximal, an operator for $\zeta$ can be defined by appealing to the spectral theorem. In either case, the operator $A^\zeta$ corresponding to $\zeta$ has a discrete spectrum. Let the eigenvalues be $\{u_j\}$ and let the corresponding eigenspaces be $\{V_j\}$. The vectors of these eigenspaces are defined as quantum states, and one can show that each eigenspace $V_j$ can be associated with a question `What is the value of $\zeta$?' together with a definite answer `$\zeta =u_j$'. This assumes that the set of values of $\zeta$ can be reduced to this set of eigenvalues, which I will justify as follows.
\bigskip

\textbf{Theorem A1.}

\textit{Let $\{u_i\}$ be the eigenvalues of the operator $A^\zeta$ corresponding to $\zeta$. Then it follows that $\Omega_\zeta$ is identical to this set of eigenvalues.}
\bigskip

\underline{Proof.}

Let $\{\zeta_i\}$ be the possible values of $\zeta$. From (\ref{u5}) we get
\begin{equation}
A^\zeta = \sum_i \sum_{j=j(i)}f_{\zeta}(n_j ) Q_i = \sum_i \zeta_i Q_i ,
\label{13}
\end{equation}
where $\{n_j; j=j(i)\}$ are the elements of the group $N$ such that $\zeta_i =f_\zeta (n_j)$, and
\begin{equation}
Q_i =r_i \sum_{j=j(i)}|v_{n_j}\rangle\langle v_{n_j}|
\label{14}
\end{equation}
for some constant $r_i$

Consider first the maximal case. Then by Theorem A2 below the eigenvalues of $A^\zeta$ are simple, so that we can write
\begin{equation}
A^\zeta = \sum u_i |u_i\rangle\langle u_i |,
\label{15}
\end{equation}
where $u_i$ and $|u_i\rangle$ are the different eigenvalues and orthogonal eigenvectors of $A^\zeta$. We have to prove that there is some connection between (\ref{13}) and (\ref{15}) in this case.

Assume that one value of $\zeta$, say $\zeta_1$, is an eigenvalue of $A^\zeta$. The other values of $\zeta$ are then given by $\zeta_i = g_i \zeta_1$, where $g_i$ is any member of the group $G$, which can be taken to be the cyclic group.

In (\ref{14}) we have $|v_{n_j} \rangle = W(n_j)|v_0\rangle=U(g_i)|v_0 \rangle$, which implies that $U(g_{i'})Q_i U(g_{i'})^\dagger$ for $i'\ne i$ is equal to some other $Q_{i''}$. It follows from $A^\zeta =\sum_i \zeta_i Q_i$ that 1) $U(g_{i'} )A^\zeta U(g_{i'})^\dagger = A^\zeta$, 2) If $\zeta_1 =u_1$ is an eigenvalue, then we must have that $\zeta_i =g_i u_1$ is an eigenvalue for all $i$, since a cyclic permutation of $\{u_i\}$ leaves (\ref{15}) invariant, and a cyclic permutation of $\{\zeta_i\}$ leaves (\ref{13}) invariant.

Let $I_0 =\{u_j : u_j =g\zeta_1\ \mathrm{for\ some}\ g\in G\}$. Since $G$ is transitive on $\Omega_\zeta$, it follows that $I_0 =\Omega_\zeta$.

Above, I have assumed that one value of $\zeta$, $\zeta=\zeta_0$ was an eigenvalue of $A^\zeta$. So, the conclusion so far is that if one value is an eigenvalue, then all values in $\Omega_\zeta$ are eigenvalues. Now the same arguments could have been used with respect to the operator $B=\gamma A^\zeta$ for some fixed constant $\gamma\ne 0$. For each $\gamma$ the conclusion is: Either (i) all values in $\Omega_\zeta$ are eigenvalues of $B$, or (ii) no values in
$\Omega_\zeta$ are eigenvalues of $B$.

Now go back to the general definition (\ref{u5}) of $A^\zeta$. Changing from $A^\zeta$ to $B$ here, amounts to changing $\zeta$ to $\zeta'=\gamma\zeta$. It is clear that we always can choose $\gamma$ in such a way that there is one value in $\Omega_{\zeta'}$ which equals the first eigenvalue of $B$. Thus, the conclusion (i) holds for one choice of $\gamma$. Now the change from $\zeta$ to $\zeta'$ also changes the measure $\mu$ which is involved in the definition of the operator and also in a corresponding resolution (\ref{uu}) of the identity. It is only one choice of $\gamma$, namely $\gamma=1$ which makes the resolution of the identity (\ref{uu}) valid, which is crucial for the theory. Thus, one is forced to conclude that $\gamma=1$, and that the conclusion (i) holds for this choice.

Hence $\Omega_\zeta$ is contained in the set of eigenvalues of $A^\zeta$. If there were one eigenvalue that is not contained in  $\Omega_\zeta$, one can use this eigenvalue as a basis for choosing $\gamma$ in the argument above, hence getting a contradiction. Thus, the two sets are identical.

Having proved this for a maximal accessible $\zeta$, it is clear that it also follows for a more general accessible $\lambda =f(\zeta)$, since the spectrum then is changed from $\{ \zeta_j \}$ to $\{f(\zeta_j)\}$.

\qed.
\bigskip

We also have the following:
\bigskip

\textbf{Theorem A2.} 

\textit{The accessible variable $\zeta$ is maximal if and only if each eigenspace $V_j$ of the operator $A^\zeta$ is one-dimensional.}
\bigskip

\underline{Proof.}

The assertion that there exists an eigenspace that is not one-dimensional, is equivalent with the following: Some eigenvalue $u_j$ correspond to at least two orthogonal eigenvectors $|j\rangle$ and $|i\rangle$. Based on the spectral theorem, the operator $A^\zeta$ corresponding to $\zeta$ can be written as $\sum_r u_r P_r$, where $P_r$ is the projection upon the eigenspace $V_r$. Now define a new accessible variable $\psi$ whose operator $B$ has the following properties: If $r\ne j$, the eigenvalues and eigenspaces of $B$ are equal to those of $A^\zeta$. If $r=j$, $B$ has two different eigenvalues on the two one-dimensional spaces spanned by $|j\rangle$ and $|i\rangle$, respectively, otherwise its eventual eigenvalues are equal to $u_j$ in the space $V_j$. Then $\zeta=\zeta(\psi)$, and $\psi\ne\zeta$ is inaccessible if and only if $\zeta$ is maximal accessible. This construction is impossible if and only if all eigenspaces are one-dimensional.
\qed
\bigskip

Point (i) in Theorem 3 follows from Theorem A1, and point (ii) follows from Theorem A2. In the maximal case there is a one-to-one correspondence between eigenvalues and eigenvectors. By (i), this gives point (iii). Point (iv) follows since $\zeta\le\eta$ in the partial ordering for some maximal accessible variable $\eta$.

\end{document}